\documentclass[final,twocolumn]{elsarticle}
\usepackage{amssymb}
\UseRawInputEncoding

\begin{document}
\def\bfe{\mathbf{e}}
\def\bfx{\mathbf{x}}
\def\tr{\mathrm{tr}}

\begin{frontmatter}

\title{First-order and pseudo-first-order transition in the high dimensional $O(N)\otimes O(M)$ model}

\author{A.O. Sorokin}
\ead{aosorokin@gmail.com}

\address{Petersburg Nuclear Physics Institute, NRC Kurchatov Institute, 188300 Orlova Roscha, Gatchina, Russia}

\begin{abstract}
Using the renormalization group approach, we consider the $O(N)\otimes O(M)$ model in four and more dimensions. We find that independently on $N$ and $M$, for $N\geq M\geq 2$, a transition can be of both the first and second order. In $d>4$, we also cannot exclude a pseudo-first-order behavior. As specific physically interesting cases, we consider the lattice version of the $O(2)\otimes O(2)$, $O(3)\otimes O(2)$ and $O(3)\otimes O(3)$ sigma models on a four dimensional hypercubic lattice. In all these cases, we find a distinct first-order transition.
\end{abstract}


\end{frontmatter}

The distinction between first-order and continuous phase transitions is obvious and consists in the presence (or absence) of a jump in an order parameter $p\in G/H$ and the internal energy. The phenomenological Landau theory formulates conditions based on group theoretical properties of an order parameter when a transition can be continuous \cite{Landau37,Lifshitz41}. The most famous condition is that the symmetric part of the cubic term of an order parameter should not contain the unit representation of the full symmetry group $G$. However, accounting for critical fluctuations, usually based on the renormalization group (RG) approach, makes these conditions necessary but not sufficient. In particular, in three dimensions $d=3$ (and generally, in $2<d<4$), a continuous transition corresponds to an attractive (stable) fixed point of RG equations, but such a point may be absent. In this case, one commonly says about a fluctuation-induced first-order transition. At first time \cite{Halperin74} (see also \cite{Coleman73}), it has been observed in the Abelian Higgs model \cite{Ginzburg50}.

In four dimensions, the fluctuational theory of phase transitions predicts that a continuous transition is described by the Gaussian fixed point with distinctive critical exponents independent on group theoretical properties of an order parameter. These exponents are perturbed by logarithmical corrections arising in the upper critical dimension $d=4$. Below four dimensions, a critical point of a continuous transition may correspond to a non-trivial (not free) conformal field theory \cite{Polyakov70}, but in $d\geq4$ the theory of critical phenomena, at least without supersymmetry, predicts only free theories. There are rigorous proofs of the triviality of the $\varphi^4$ theory for $d>4$ \cite{Aizeman81,Aizeman82,Frolich82}, while for $d=4$ such a proof is not exist, and the role of perturbative logarithmic corrections is discussed (see, e.g. \cite{Frolich83,Kenna93,Kenna04,Stevenson05,Balog06,Lundow09,Lundow11,Akiyama19}).

Recently \cite{Akiyama19}, using the specific non-perturbative RG approach (so-called higher-order tensor RG) it has been found the weak first-order transition in the four-dimensional Ising model. The hidden heat of the transition is so small that it is necessary to consider huge size lattices (up to $1024^4$) to discover it. This makes it very difficult to check the result with other methods, for example with Monte Carlo simulations.

Offhand, one can propose two ways to explain this result. The first way is that a transition is of pseudo-first order, i.e. the sings of a first-order transition (e.g., a jump in the internal energy) are observed only on finite-size lattices and disappear in the thermodynamic limit.  In terms of RG, the pseudo-first order behavior corresponds to the situation when a RG-trajectory starting in the stability region (where the potential is stable) passes through the region boundary but tends to the fixed point locating inside the region or on the boundary. To stabilize the potential, one should add next terms resolved by symmetry in the order parameter expansion, like $\varphi^6$. It is these terms (in accordance with the Landau theory) that correspond to the appearance of the observed jump in the internal energy on finite-size lattices. But actually, such a situation cannot be realized in the four-dimensional $O(N)$ model (or other models with unique $\varphi^4$ term), where the boundary of the stability region is a single point, namely the Gaussian fixed point.

If the transition is indeed of the first order, then one can expect the presence of any non-perturbative effects (remind, it is believed that perturbative corrections keep the theory trivial). The most obvious non-perturbative aspect of the Ising model is that fluctuations are topological defects, domain walls. Of cause, the presence of topological defects in itself does not guarantee the appearance of new effects in the critical behavior. E.g., monopole-like defects in the three-dimensional $O(3)$ model are not relevant \cite{Cardy80,Antunes02}, and in general, the 3D $O(N)$ model is successfully described by perturbative approaches, although it can contains topological defects of any types. Nevertheless, topological defects may affect the critical behavior, and to test such a possibility it is useful to consider at least two cases: a system with the order parameter space $G/H=\mathbb{Z}_2\otimes \mathcal{M}$, where $\mathcal{M}$ is a connected homogeneous space, and a system with topological defects of another type affecting the critical behavior. Both cases are present in the $O(N)\otimes O(M)$ model.

The order parameter space in this model is a Stiefel manifold $V_{N,M}\equiv O(N)/O(N-M)$ corresponding to a set of orientations of $M$ $N$-dimensional vectors. In the particular case $M=N$, the order parameter contains the discrete part equivalent to the parameter of the Ising model $G/H=O(N)\equiv\mathbb{Z}_2\otimes SO(N)$, so one should expect that a transition is of the first order. Another interesting feature of this model is that in the cases $N=M$ and $N=M+1$ the fundamental group is non-trivial $\pi_1(V_{N,M})=\mathbb{Z}_2$ for $M>1$, and so-called $\mathbb{Z}_2$-vortices are present. Remind that usual vortices play a crucial role in the critical behavior of systems from the universality class of the $O(2)$ model. In this case, perturbative fluctuations can be linearized (by the Berezinskii -- Villain transformation \cite{Berezinsky70,Villain75,Kadanoff77}) and integrated out, so the resulting system has a transition from the same universality class. Of course, this is due to the fact that $SO(2)$ is abelian.

The influence of non-abelian $\mathbb{Z}_2$-vortices on the thermal and critical behavior is observable in all dimensions $2\leq d<4$. Thus, in two dimensions these defects lead to a rather sharp change in the  behavior of the $V_{3,2}$ model from low-temperature, describing by the $O(4)$ sigma model, to some high-temperature \cite{Southern95,Azaria01,Hasselman14,Sorokin19}. In the $V_{3,3}$ model, vortices changes an Ising-like transition to a first-order one \cite{Sorokin17,Sorokin18}. In $2+\epsilon$ dimensions, a transition of the $V_{3,2}$ model belongs to the class universality of the $O(4)$ model \cite{Azaria90,Azaria93,Pelissetto01}, but close to $\epsilon\approx1$ a transition becomes of the first order \cite{Zumbach95}. In three dimensions, the $V_{N,N}$ model has a distinct first-order transition, but the situation for the $V_{3,2}$ model is still controversial. So, the $4-\varepsilon$ \cite{Kawamura90,Sokolov95,Calabrese04,Sokolov20}, $1/N$ \cite{Gracey02,Gracey02-2} expansions and the non-perturbative RG \cite{Zumbach93,Zumbach94,Zumbach94-2,Delamotte00,Delamotte03,Delamotte16} predict the first order, while the perturbative (fixed-dimensional) RG \cite{Sokolov94,Pelissetto01-2,Sokolov02,Sokolov03,Parruccini03,Pelissetto04,Delamotte10} and the conformal bootstrap \cite{Nakayama14,Nakayama15,Henriksson20} declare a second-order transition.

Monte Carlo simulations for three-dimensional models from the universality class of the $V_{3,2}$ model also give controversial results (a first-order transition in a Heisenberg antiferromagnet on a stacked-triangular lattice \cite{Diep08} and helimagnets \cite{Sorokin14}, see however the recent work \cite{Kawamura19}). Nevertheless, direct simulations of the lattice version of the $O(N)\otimes O(M)$ sigma model predict a distinct first-order transition for the $V_{2,2}$ \cite{Kunz93,Loison98}, $V_{3,2}$ \cite{Loison99} and $V_{3,3}$ \cite{Diep94,Loison00} cases. These results are not a defect of the model. For $N>M+2$, the model has a second-order or a weak first-order transition with critical exponents consistent with theoretical predictions \cite{SorokinFut}. Moreover, although the two-dimensional $V_{3,3}$ model also has a first-order transition, but, e.g., the 2D $V_{2,2}$ model shows more difficult behavior, namely when a Berezinskii-Kosterlitz-Thouless transition and an Ising one occur at different temperatures \cite{Sorokin19,Sorokin18}.

\begin{figure}[t]
    \center
    \includegraphics[scale=0.25]{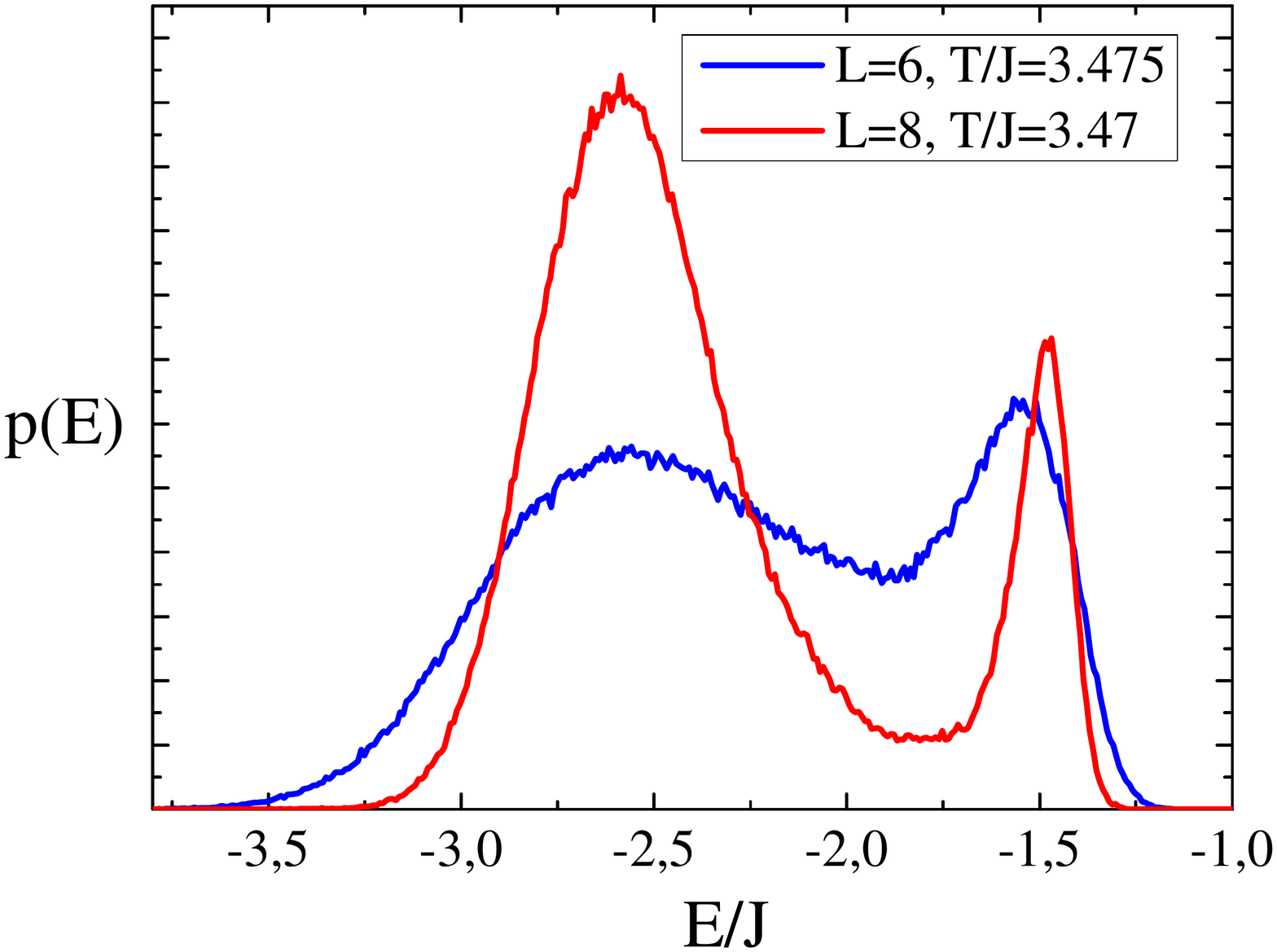}%
    \caption{\label{fig1} Internal energy distribution in the $O(2)\otimes O(2)$ model}
\end{figure}%
\begin{figure}[t]
    \center
    \includegraphics[scale=0.25]{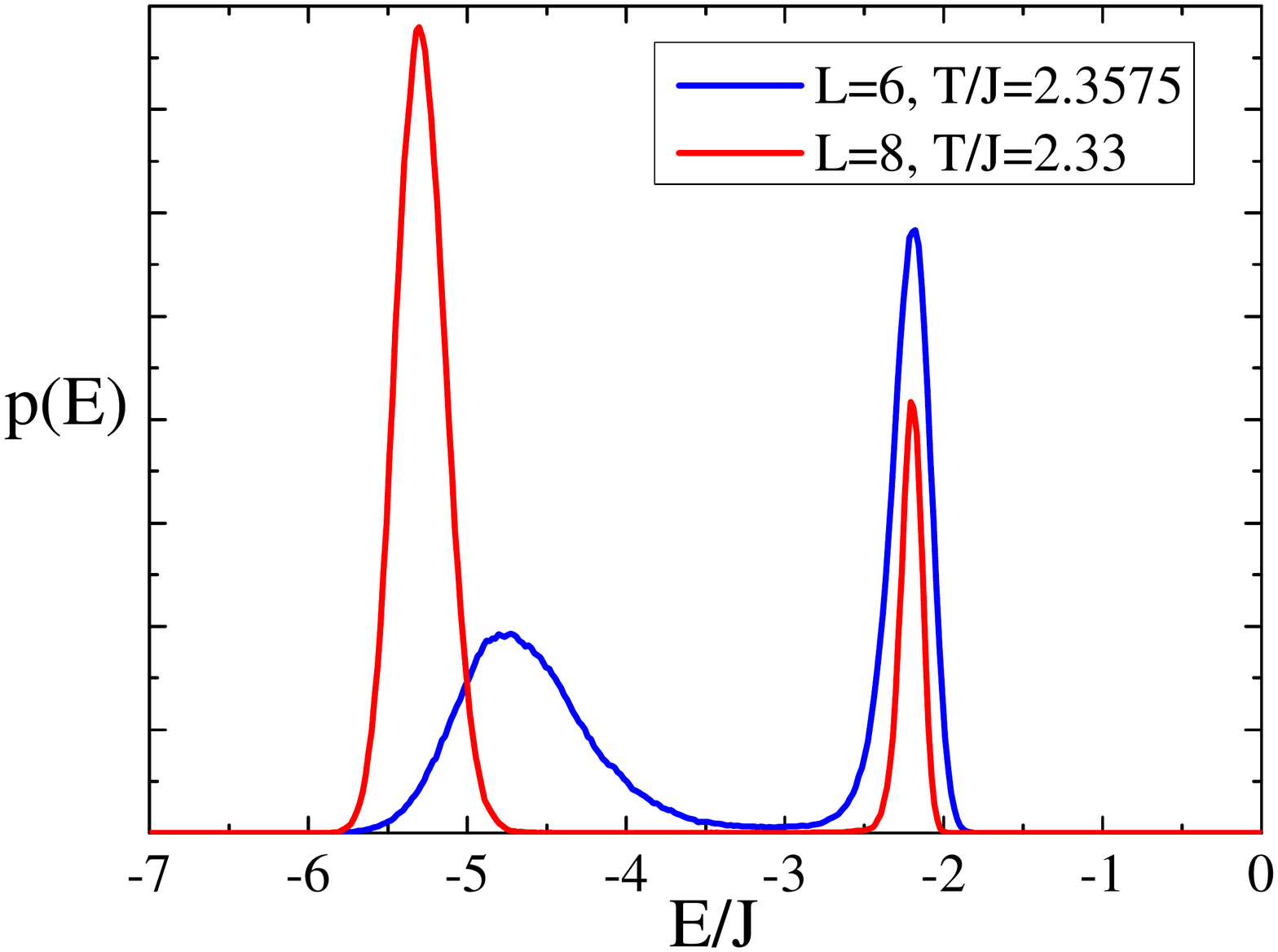}%
    \caption{\label{fig2} Internal energy distribution in the $O(3)\otimes O(3)$ model}
\end{figure}%
\begin{figure}[t]
    \center
    \includegraphics[scale=0.25]{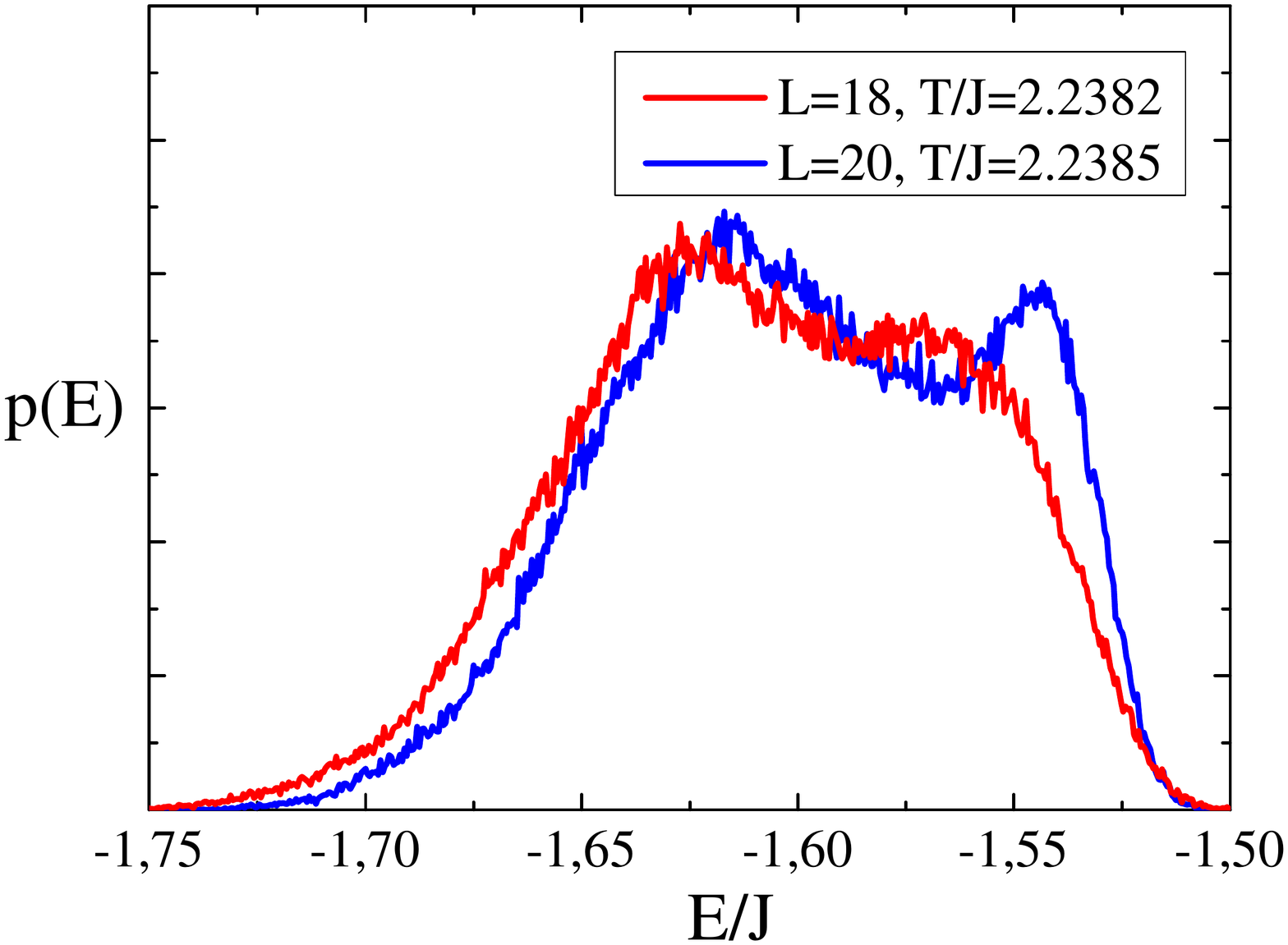}%
    \caption{\label{fig3} Internal energy distribution in the $O(3)\otimes O(2)$ model}
\end{figure}%
Somewhat surprisingly, the $V_{2,2}$, $V_{3,2}$ and $V_{3,3}$ models exhibit the distinct first-order behavior even in four dimensions. To obtain this, we perform Monte Carlo simulations of the corresponding lattice models. Generally, the lattice version of the $O(N)\otimes O(M)$ sigma model is described by the Hamiltonian \cite{Kunz93}
\begin{equation}
    H=-J\sum_{\bfx,\mu}\tr\,\Phi_\bfx^T\Phi_{\bfx+\bfe_\mu},\quad \mu=1,\ldots,4,
    \label{lattice-model}
\end{equation}
where $\bfe_\mu$ is a unit vector of a hypercubic lattice, $J>0$, $\Phi$ is a $N\times M$ matrix composed of $M$ unit mutually orthogonal $N$-vectors. We use the Wollf cluster algorithm \cite{Wollf89}. In figs. \ref{fig1} -- \ref{fig3}, we see a double-peak structure of distributions for the internal energy. Such a structure is typical for a discontinues transition.

Although the initial motivation for this work is to find any non-perturbative effects leading to a change of the critical behavior predicted by the RG approach, it turns out that in the case of the $O(N)\otimes O(M)$ model, the results obtained by Monte Carlo simulations and described above can be easily confirmed using the perturbative RG.

The corresponding Ginzburg -- Landau functional for the model (\ref{lattice-model}) reads \cite{Kawamura90}
$$
    F=\int d^dx\times
$$
$$
    \times\left(\frac12\sum_{n=1}^{M}\bigl((\partial_\mu\mathbf{\phi}_n)^2+r\mathbf{\phi}_n^2\bigr)+
    \frac u{4!}\left(\sum_{n=1}^M\mathbf{\phi}_n^2\right)^2+\right.
$$
\begin{equation}
    \left.
    +\frac v{4!}\sum_{n,m=1}^M\left((\mathbf{\phi}_n\mathbf{\phi}_m)^2-\mathbf{\phi}_n^2\mathbf{\phi}_m^2\right)\right),
    \label{GLW-model}
\end{equation}
where $\phi_n$ is a $N$-component vector field, $\Phi=(\phi_1,\ldots,\phi_M)$. The 1-loop beta-functions in the $4-\varepsilon$ expansion with the $\overline{\mathrm{MS}}$ scheme after rescaling are
\begin{equation}
    \beta_u=-\varepsilon u+\frac16(NM+8)u^2-\frac16(N-1)(M-1)v(2u-v),
\end{equation}
\begin{equation}
    \beta_v=-\varepsilon v+2uv+\frac16(M+N-8)v^2.
\end{equation}
In this scheme, the series for the beta-functions do not contain poles of $\varepsilon$, so one can extrapolate this result to $\varepsilon\to0$.

\begin{figure}[t]
    \center
    \includegraphics[scale=0.75]{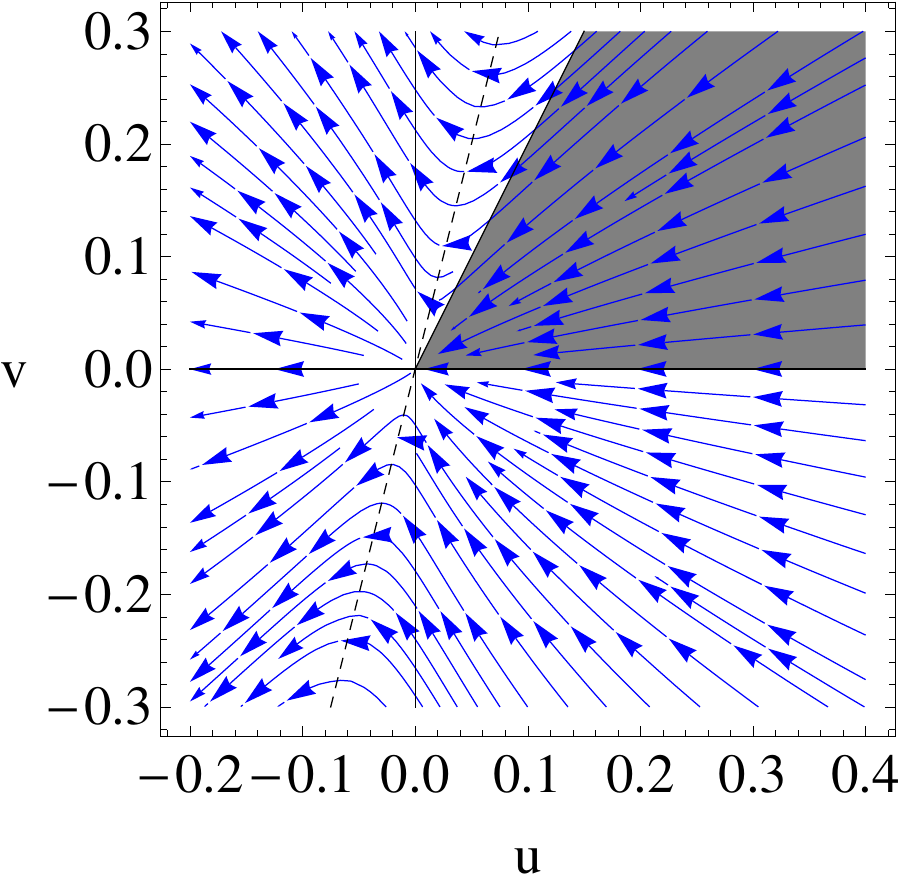}%
    \caption{\label{fig4} RG-diagram of the 4D $O(3)\otimes O(2)$ model}
\end{figure}%
Fig. \ref{fig4} shows the RG-diagram of the $O(3)\otimes O(2)$ model. The gray sector corresponds to the stability of the potential (\ref{GLW-model})
\begin{equation}
    u>0,\quad v>0,\quad \frac{M}{M-1}u-v>0.
\end{equation}
We show only this case, the picture remains qualitatively the same for all $N\geq M\geq2$. One can see and it can be easily proved that the RG-flow does not pass through the line $v=0$, but the flow passes through another boundary of the stability region $v=\frac{M}{M-1}u$ with the flow velocity in the direction perpendicular to this boundary having the value (independently on $N$ and $\varepsilon$)
\begin{equation}
    v_{\mathrm{flow}}=\frac{M+2}{3(M-1)}u^2.
\end{equation}
In other words, for all $N\geq M\geq2$ there are RG-trajectories starting from the stability region but which are not attracted by the Gaussian fixed point $u=v=0$ and pass through the stability region boundary. Thus, in the $O(N)\otimes O(M)$ model a first-order transition induced by perturbative fluctuations can occur. We emphasize that this result does not depend on the topological or other properties of the order parameter space and does not require any additional non-perturbative effects.

\begin{figure}[t]
    \center
    \includegraphics[scale=0.75]{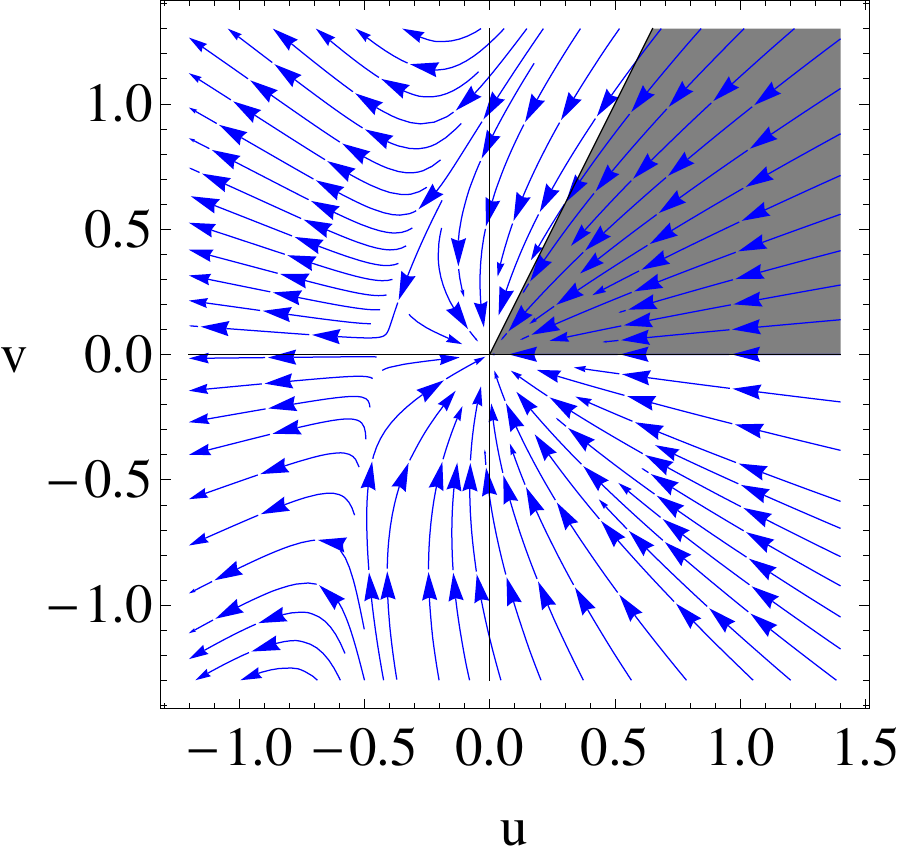}%
    \caption{\label{fig5} RG-diagram of the 5D $O(3)\otimes O(2)$ model}
\end{figure}%
The naive extrapolation of the beta-function series to the region of negative $\varepsilon$ shows that some of trajectories, starting from the stability region and tending to the Gaussian fixed point, can leave this region (see fig. \ref{fig5}). It corresponds to a pseudo-first-order transition. So, we cannot exclude such a behavior in $d>4$. Such a possibility for the five-dimensional Ising model has been discussed in \cite{Lundow11}.

In conclusion, we note that the independence of the result on $N$ and $M$ allow to suppose that the accounting of next orders of the perturbative expansion does not change the result. However, we do not exclude of non-perturbative effects. So, e.g., the result predicts that some trajectories in the $N=M$ case attract to the Gaussian fixed point. It means that some systems belonging to the universality class of the $O(N)\otimes O(N)$ model may have a second-order transition. Taking into account the result of \cite{Akiyama19},  we expect that the same effects, which change the order of a transition in the Ising model, are also relevant in the $N=M$ case. To investigate this, one should find suitable reliable methods.

\medskip

This work was supported by the Theoretical Physics and Mathematics Advancement Foundation 'BASIS' (project No. 19-1-3-38-1).

\end{document}